\documentclass[twoside,a4paper,english]{IEEEtran}

\usepackage[dvips]{graphicx}
\usepackage{psfrag}
\usepackage[T1]{fontenc}
\usepackage[latin9]{inputenc}
\usepackage{float}
\usepackage{amsthm}
\usepackage{amsmath}
\usepackage{amssymb}
\usepackage{graphicx}
\usepackage{esint}
\makeatletter

\theoremstyle{plain}

\usepackage{amsthm}

\IEEEoverridecommandlockouts
\@ifundefined{definecolor}{\usepackage{color}}{}
\usepackage{algorithmic}
\usepackage{algorithm}

\makeatother

\usepackage{babel}
\providecommand{\theoremname}{Theorem}

\begin{document}

\title{Spectrum Monitoring Using Energy Ratio Algorithm For OFDM-Based Cognitive Radio Networks}
\author{{\normalsize {Abdelmohsen Ali,~\emph{Student Member,~IEEE} and Walaa Hamouda,~\emph{Senior Member,~IEEE}}\thanks{This research was
supported by the Natural Sciences and Engineering Research Council of Canada (NSERC) Grants N008861 and N01268.} \thanks{
The authors are with the Department of Electrical and Computer Engineering, Concordia University, Montreal, Quebec, H3G 1M8, Canada
(e-mail:(ali\_abde,hamouda)@ece.concordia.ca). } }}

\maketitle

\thispagestyle{empty}
\begin{abstract}

This paper presents a spectrum monitoring algorithm for Orthogonal Frequency Division Multiplexing (OFDM) based cognitive radios by which the primary user reappearance can be detected during the secondary user transmission. The proposed technique reduces the frequency with which spectrum sensing must be performed and greatly decreases the elapsed time between the start of a primary transmission and its detection by the secondary network. This is done by sensing the change in signal strength over a number of reserved OFDM sub-carriers so that the reappearance of the primary user is quickly detected. Moreover, the OFDM impairments such as power leakage, Narrow Band Interference (NBI), and Inter-Carrier Interference (ICI) are investigated and their impact on the proposed technique is studied. Both analysis and simulation show that the \emph{energy ratio} algorithm can effectively and accurately detect the appearance of the primary user. Furthermore, our method achieves high immunity to frequency-selective fading channels for both single and multiple receive antenna systems, with a complexity that is approximately twice that of a conventional energy detector.

\end{abstract}



\section{Introduction}

Nowadays, static spectrum access is the main policy for wireless communications. Under this policy, fixed channels are assigned to licensed users or primary users (PUs) for exclusive use while unlicensed users or secondary users (SUs) are prohibited from accessing those channels even when they are unoccupied. The idea of a cognitive radio (CR) was proposed in order to achieve more efficient utilization of the RF spectrum~\cite{ref:BasicCRIdea}. One of the main approaches utilized by cognitive networks is the interweave network model~\cite{ref:interweaveSensingConcept} in which SUs seek to opportunistically use the spectrum when the PUs are idle. Primary and secondary users are not allowed to operate simultaneously. In this method, secondary users must sense the spectrum to detect whether it is available or not prior to communication. If the PU is idle, the SU can then use the spectrum, but it must be able to detect very weak signals from the primary user by monitoring the shared band in order to quickly vacate the occupied spectrum. During this process, the CR system may spend a long time, known as the sensing interval, during which the secondary transmitters are silent while the frequency band is sensed. Since the CR users do not utilize the spectrum during the detection time, these periods are also called quiet periods (QPs)~\cite{ref:quietperiodmangforCR}. In the IEEE 802.22 system, a quiet period consists of a series of consecutive spectrum sensing intervals using energy detection to determine if the signal level is higher than a predefined value, which indicates a non-zero probability of primary user transmission. The energy detection is followed by feature detection to distinguish whether the source of energy is a primary user or noise~\cite{ref:quietperiodadaptivescheme}~\cite{ref:energyandfeaturedetforquietperiods}. This mechanism is repeated periodically to monitor the spectrum. Once the PU is detected, the SU abandons the spectrum for a finite period and chooses another valid spectrum band in the spectrum pool for communication.

If the SU must periodically stop communicating in order to detect the emergence of the PU, two important effects should be studied. (1) During quite periods, the SU receiver may lose its synchronization to the SU transmitter which causes an overall degradation in the secondary network performance. This is a problem when the underlying communication technique is sensitive to synchronization errors as in OFDM~\cite{ref:QPsynchproblem}. (2) The throughput of the secondary network during sensing intervals is reduced to zero which degrades the Quality of Service (QoS) for those real-time applications like Voice over IP (VoIP)~\cite{ref:IEEE80222refnumbers}. The impact becomes more severe if the duration of the sensing intervals is too large as the average throughput of the secondary network becomes very low. On the other hand, if this duration is too small, then the interference to the primary users is increased since spectrum sensing provides no information about the frequency band of interest between consecutive sensing intervals.

In this area, there have been research efforts which attempt to minimize the time duration for spectrum monitoring by jointly optimizing the sensing time with the detection threshold~\cite{ref:effspectrumMonitoring}. The PU throughput statistics are considered to protect the PU while the sensing time is minimized. In conventional systems, traditional spectrum sensing is applied once before the SU communication and is not be repeated again unless the monitoring algorithm indicates that a primary signal may be present in the band. If monitoring determines correctly that there is no primary signal in the band, then the time that would have been spent performing spectrum sensing is used to deliver packets in the secondary network. Therefore the spectrum efficiency of the secondary network is improved. If spectrum monitoring detects a primary signal in the band during a time period in which spectrum sensing would not have been scheduled, then the disruption to the primary user can be terminated more quickly and hence the impact of secondary communications on the primary user is reduced. Based on this description, the SU receiver should follow two consecutive phases, namely sensing phase and monitoring phase, where the former is applied once for a predefined period.

Yet, another approach is utilized by~\cite{ref:MonitoringDuringReception} where the spectrum is monitored by the CR receiver during reception and without any quiet periods. The idea is to compare the bit error count, that is produced by a strong channel code like a Low Density Parity Check (LDPC) code, for each received packet to a threshold value. If the number of detected errors is above certain value, the monitoring algorithm indicates that the primary user is active. The threshold is obtained by considering the hypothesis test for the receiver statistics when the primary signal is absent and the receiver statistics for the desired Secondary-to-Primary power Ratio (SPR). Although this technique is simple and adds almost no complexity to the system, the receiver statistics are subject to change by varying the system operating conditions. In real systems, there are many parameters that can affect the receiver error count such as RF impairments including Phase Noise (PN) and Carrier Frequency Offset (CFO), Sampling Frequency Offset (SFO), and NBI. The error count will depend not only on the presence of a primary signal but it will also depend on the characteristics of those impairments. Also, the receiver statistics may change from one receiver to the other based on the residual errors generated from estimating and compensating for different impairments. Since it is difficult to characterize the receiver statistics for all CR receivers, it is better to devise an algorithm that is robust to synchronization errors and channel effects.

OFDM is a multi-carrier modulation technique that is used in many wireless systems and proven as a reliable and effective transmission method. For these reasons, OFDM is utilized as the physical layer modulation technique for many wireless systems including DVB-T/T2, LTE, IEEE 802.16d/e, and IEEE 802.11a/g. Similar to other wireless networks, OFDM is preferred for cognitive networks and has been already in use for the current cognitive standard IEEE 802.22. On the other hand, OFDM systems have their own challenges that need special treatment~\cite{ref:OFDMandCR}. These challenges include its sensitivity to frequency errors and the large dynamic range of the time domain signal. Moreover, the finite time-window in the receiver DFT results in a spectral leakage from any in-band and narrow band signal onto all OFDM sub-carriers.

The traditional spectrum monitoring techniques, that rely on the periodic spectrum sensing during quiet periods, apply their processing over the received time domain samples to explore a specific feature to the primary user. Further, it is totally encouraged to remove the quiet periods during the monitoring phase in order to improve the network throughput. In fact, the signal construction for the secondary user can assist the spectrum monitoring to happen without involving QPs. When the secondary user utilizes OFDM as the physical transmission technique, a frequency domain based approach can be employed to monitor the spectrum during the CR reception only if the SU transmitter adds an additional feature to the ordinary OFDM signal. In this paper, we propose a spectrum monitoring technique, namely the \emph{energy ratio} (ER) technique, that is suitable for OFDM-based cognitive radios. Here, the transmitter helps this frequency domain based spectrum monitoring approach by introducing scheduled null-tones by which the spectrum can be monitored during CR reception. This monitoring technique is designed to detect the reappearance of the primary user which also uses OFDM technique. Here, different signal chain impairments due to CFO, SFO, and NBI as well as frequency selective fading channels are considered. The technique operates over the OFDM signal chain and hence, it does not require to wait for the decoded bits. This implies fast response to PU appearance. Furthermore, the most important OFDM challenges for cognitive radios like power leakage are investigated and their effects on the proposed monitoring technique are considered.

The paper is organized as follows. Section~\ref{sec:systemmodel} summarizes the overall system model. In section~\ref{sec:energyratioalgorithm}, the \emph{energy ratio} technique is discussed. In addition, we present performance analysis for AWGN channels under perfect synchronization and neglecting power leakage in section~\ref{sec:ERanalysis}. OFDM challenges and non-perfect synchronization environment are considered in section~\ref{sec:OFDMchallenges}. Further, we extend the analysis and study to frequency selective fading channels and multi-antenna systems in section~\ref{sec:channelMIMO}. The complexity of the \emph{energy ratio} is analyzed and an architecture is also proposed in~\ref{sec:complexity}. Finally, the performance is evaluated with computer simulations in section~\ref{sec:Results}.

\section{System Model}
\label{sec:systemmodel}

The secondary user physical layer model is designed in order to investigate and verify our spectrum monitoring algorithm. This model is very close to the OFDM system followed by~\cite{ref:OFDMandCR}. At the transmitter side, data coming from the source is firstly segmented into blocks where each block is randomized, channel encoded, and interleaved separately. After interleaving, the data is modulated by the constellation mapper. The frequency domain OFDM frame is constructed by combining: (a) One or more training symbols or preambles that are used for both time and frequency synchronization at the receiver side. (b) The modulated data. (c) The BPSK modulated pilots which are used for data-aided synchronization algorithms employed by the receiver. Each $N_s$ encoded complex data symbols generated by the frame builder are used to construct one OFDM symbol by employing the IDFT block that is used to synthesize the OFDM symbol, where $N_s$ denotes the number of sub-carriers per one OFDM symbol. Thus, the $n^\mathrm{{th}}$ time-domain sample of the $m^\mathrm{{th}}$ symbol can be expressed as given by (\ref{equ:OFDMTDsignal}) where $C(k,m)$ is the modulated data to be transmitted on the $m^\mathrm{{th}}$ OFDM symbol with the $k^\mathrm{{th}}$ sub-carrier.
\begin{eqnarray}
s(n, m)&=&\frac{1}{\sqrt{N_s}} \sum_{k=-N_s/2}^{N_s/2-1} C(k, m)\,e^{j2\pi kn/N_s} \label{equ:OFDMTDsignal}
\end{eqnarray}

To reduce the effect of Inter-Symbol Interference (ISI), the last $N_g$ samples of the time domain OFDM symbol are copied to the beginning of the symbol in order to form a guard time or cyclic prefix. Therefore, the OFDM block has a period of $T_s=(N_s+N_g)/F_s$ where $F_s$ is the sampling frequency. At the receiver, the inverse blocks are applied. After timing synchronization (frame detection, start of symbol timing, and SFO estimation and compensation) and frequency synchronization (CFO estimation and correction), the cyclic prefix is removed. Then, the received OFDM symbol is transformed again into the frequency domain through an $N_s$ point DFT. The channel is then estimated and the received data is equalized. The complex data output is then mapped to bits again through the De-mapper. De-interleaving, decoding, and De-randomization are applied later to the received block to recover the original source bits.

\begin{figure}[t]
\centering
\includegraphics[scale=0.39]{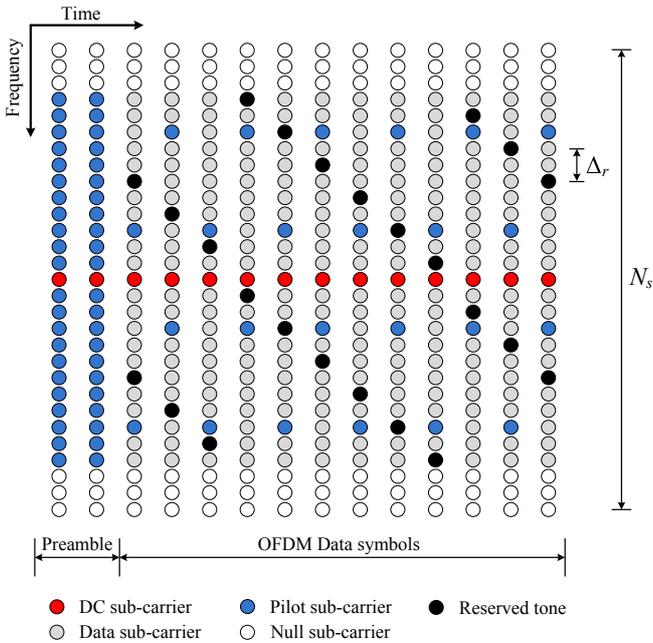}
\caption{Time-Frequency allocation for one OFDM frame to explore different sub-carrier types}
\label{fig:OFDMFrameStructureReservedTones}
\end{figure}

From the network point of view, we consider a cognitive radio network of $K$ SUs and one $PU$. The PU occupies a spectrum of a certain bandwidth for its transmission, while the same spectrum is shared by the SUs. In fact, the spectrum is totally utilized by one SU (the master node or the fusion node) to send different data to the other $K-1$ SUs (the slave nodes). This model was originally introduced for Frequency Division Multiple Access (FDMA)~\cite{ref:modelrefsigprocessing} but it has been modified later to suite the OFDM environment. Currently, this model particularly matches two promising solutions, namely Ecma-392 and IEEE 802.11af, that employ OFDM as the underlying physical transmission technique~\cite{ref:Ecma80211afcompare}. The standards introduce cognitive radio approach to the TV white space. Usually, a secondary user should get necessary information from TV white space database which maintains a list of the unused TV channels geometrically. However, the standards specify channel power management functionality in order to update the available channel lists.

The current Ecma-392 standard supports spectrum sensing functionality to periodically check the existence of incumbent signals on the current operating channel. Ecma-392 has specified the operation in only single TV channel which can be one of three channel bandwidths of 6 MHz, 7 MHz, or 8 MHz according to regulatory domain. The objective is that the secondary user can utilize the full band on which the primary user operates. The IEEE 802.11af standard is an extension to the Wireless Local Area Network (WLAN). The channel bandwidths in this standard can be adaptively changed when several adjacent TV channels are available. Again, the fusion node (the access point) utilizes the whole primary user band to broadcast the downlink signal to all slaves. In reality, our algorithm is demonstrated by a more general model which does not perfectly match the implementation of  either Ecma-392 or IEEE 802.11af. The main difference is that our model has no limitations on the channel bandwidth, the channel characteristics, or even the frequency tolerances.

In our model, the fusion node constructs OFDM frames in the downlink path such that the same pilots are transmitted to all slaves but the data sub-carriers are allocated in time and frequency for different users based on a predefined scheduling technique. For the return path, Orthogonal Frequency Division Multiple Access (OFDMA) is assumed to divide the spectrum and the time into distinct and non-overlapping channels for different slaves, so that interferences between the slaves is avoided. The fusion node fully controls the timing of each slave, possibly by letting the slave know the required time advance or delay, so that the combined signal from all slaves seem to be synchronized at the fusion node receiver. In this case, the fusion node can convert the signal back to the frequency domain in order to extract the data and control information from different slaves. A valid assumption is that the slaves can send important information such as spectrum monitoring decisions and channel state information over a logical control channel in the return path. The master node can simply apply a majority rule based on the received monitoring decisions to decide whether to stop transmission or not.

\section{Energy Ratio Algorithm}
\label{sec:energyratioalgorithm}
On the time-frequency grid of the OFDM frame and before the IDFT, a number of tones, $N_{RT}$, are reserved for the spectrum monitoring purposes. These tones are reserved for the whole time except the time of the training symbol(s) in order not to change the preamble waveform, which is used for synchronization at the receiver. The proposed OFDM frame is shown in Fig.~\ref{fig:OFDMFrameStructureReservedTones}. Notice that we allocate the reserved tones dynamically so that their indices span the whole band when successive OFDM symbols are considered in time. The tones are advanced by $\Delta_r$ positions every OFDM symbol. When the last index of the available sub-carriers is reached, the spanning starts again from the first sub-carrier. Hence, by considering small values for $\Delta_r$, the reserved tone sequence injected to the energy ratio spans the whole band. The reasons for this scheduling are:  (1) The primary user may have some spectrum holes because of using OFDM as well. If the reserved tones from the SU are synchronized with those spectrum holes in the PU side, then the algorithm will fail. On the contrary, if the PU uses a traditional single carrier modulation technique like QAM, this issue does not have a harm effect on the algorithm since the PU signal has a flat spectrum over the entire band. (2) The reserved tones typically occupy narrow band and the primary to secondary channel may introduce notch characteristics to this narrow band resulting in detecting lower primary power level, which is referred to the narrow band problem. Therefore, it is recommended that the reserved tones are rescheduled by changing the value of $\Delta_r$ over time to mitigate the channel effect and to protect the reserved tones from falling into primary holes. Of course, all SUs should know the code for this scheduling in prior.

\begin{figure}[t]
\centering
\includegraphics[scale=0.39]{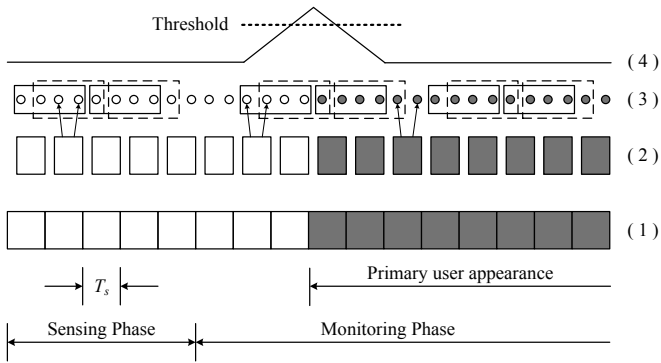}
\caption{Energy Ratio processing details. (1) The time domain sequence for the OFDM blocks. (2) Frequency domain samples. (3) Reserved tones processing with two sliding windows for $N_{RT}=2$ and $N=4$. (4) Decision making variable, $X_k$.}
\label{fig:EnergyRatioWithTime}
\end{figure}

Based on the signal on the reserved tones at the receiver, the secondary user can monitor the band and test the primary user appearance. In fact, the traditional radiometer may be employed to measure the primary signal power and the secondary noise power by accumulating the energy of those reserved tones. As a consequence, the primary signal power can be detected if this energy exceeds a predefined threshold. However, this approach does not necessary guarantee the primary user detection as the spectral leakage of the neighbouring sub-carriers will affect the energy at the reserved tones even for no in-band primary signal~\cite{ref:spectrumMonitoringFilterBank}. Here, we propose another decision making criterion that has a powerful immunity for this power leakage. In fact, the power leakage, the ICI resulted from the residual CFO and SFO errors, and even the effect of NBI can be overcome by our approach.

The overall algorithm is illustrated by Fig.~\ref{fig:EnergyRatioWithTime}. It is assumed that the primary signal appears after some time during the monitoring phase. At the secondary receiver, after CP removal and frequency domain processing on the received signal, the reserved tones from different OFDM symbols are combined to form one sequence of complex samples. Two consecutive equal-sized sliding windows are passed over the reserved tone sequence in the time direction. The energy of the samples that fall in one window is evaluated and the ratio of the two energies is taken as the decision making variable and hence the name \emph{energy ratio}.

The algorithm aims to check the change in variance on the reserved tones over time. In a mathematical form, let $Z_i$ be the $i^{\mathrm{th}}$ sample of the reserved tone sequence. The decision making variable, $X_k$, can be defined as given by (\ref{equ:EnergyRatio}) where $N$ is the number of samples per window, $U_k$ is the energy of the second window, $V_k$ is the energy of the first window, and $k$ is an integer such that $k\,=\,1,\,2,\,3,\,...$

\begin{eqnarray}
X_k&=&\frac{U_k}{V_k}\,=\,\frac{\sum_{i=N+k}^{2N+k-1} \big| Z_i \big| ^2}{\sum_{i=k}^{k+N-1} \big| Z_i \big| ^2} \label{equ:EnergyRatio}
\end{eqnarray}

It should be mentioned that the reserved tones processing done by the \emph{energy ratio} algorithm starts from the beginning of the sensing phase. Meaning that, the decision making variable is evaluated during both sensing and monitoring phases. However, it provides decisions only during monitoring phase. During the sensing phase, if the decision from the spectrum sensing algorithm is that the PU is inactive, then the \emph{energy ratio} algorithm has been properly calibrated to be able to detect the appearance of the PU during monitoring phase. Calibration means that both sliding windows are filled with pure unwanted signals. During the monitoring phase, the receiver monitors the reserved tones by evaluating the parameter, $X_k$. If it exceeds a certain threshold, then the secondary user assumes that there is a power change on the reserved tones which perhaps due to the primary user appearance and it is time to vacate the band. If not, the secondary user can continue transmission. Indeed, if there is no primary user in band, then the energy of each window still involves only the strength of the unwanted signals including the noise, the leakage from the neighbouring sub-carriers, and the effects of ICI produced by the residual synchronization errors. Therefore, if $N$ is large enough, the ratio will be very close to unity since the strength of the unwanted signals does not offer significant changes over time.

Once the primary user appears, the second window will have two types of signalling which are the primary user interference and the unwanted signals. Meanwhile, the first window will only maintain the unwanted signals without the primary user interference. The ratio of the two energies will result in much higher values when compared to one. The value will of course depend on the primary user power. When the two windows slide again, the primary signal plus the unwanted signals will be observed by the two windows and the decision making variable returns to the initial state in which the ratio is close to unity. Thus, we can expect that the decision variable produces a spike when the primary user is detected. Otherwise, it changes very slowly maintaining the \emph{energy ratio} close to one as shown in Fig.~\ref{fig:EnergyRatioWithTime} part (4).

This approach can resist the different impairments involved in the received signal on the account of reducing the throughput of the secondary user by the ratio of the number of reserved tones to the number of useful tones. However, this reduction can be easily overcome since OFDM systems allow adaptive modulation where good conditioned sub-carriers are loaded with higher modulation order.

For the previous discussion, it is assumed that the primary user should appear at the boundaries of the OFDM blocks. Therefore, the reserved tones should have the full power, that is supposed to be for those sub-carrier indices, of the primary user when it is active. In reality, the primary user may appear any time within any OFDM block in the monitoring phase. In this case, we have to consider two effects. (1) The FFT window applied by the SU receiver will have a time-shifted version of the PU signal which involves a phase rotation to the PU sub-carriers. Since the energy is the useful parameter for our algorithm, the phase shift is acceptable to happen with no effect on the algorithm. (2) The power on the reserved tones will not have the full power transmitted by the primary user on those sub-carriers since part of the signal is truncated. However, the next OFDM symbol will have that full power. Similar to the near-far problem, if the PU power is large enough, then the reserved tones form the first OFDM symbol, in which PU signal appears, are considered to be full. Otherwise, the reserved tones from this OFDM symbol are considered as noise if $N\gg N_{RT}$.

\begin{figure*}[!t]
\normalsize
\setcounter{equation}{7}
\begin{eqnarray}
F_{X}(x)&=& \mathrm{Prob}\big[U\leq xV\big]\,=\, \int_{0}^{\infty} \int_{0}^{xv} f_{UV}(u,v) \,\mathrm{d}u\,\mathrm{d}v \nonumber \\
&=& \int_{0}^{\infty} \int_{0}^{xv} \frac{1}{2^{2N}\,\sigma_v ^{2N}\,\sigma_u ^{2N}\,\Gamma(N)\,\Gamma(N)}\,u^{N-1}\,e^{-u/(2\sigma_u ^2)}\,v^{N-1}\,e^{-v/(2\sigma_v ^2)} \,\mathrm{d}u\,\mathrm{d}v \nonumber \\
&=& \frac{1}{2^{2N}\,\Gamma^2(N)} \int_{0}^{\infty} \Bigg[\int_{0}^{x\acute{v}\,\sigma^2_v/\sigma _u^2} \acute{u}^{N-1}\,e^{-\acute{u}}\,\mathrm{d}\acute{u}\,\Bigg]\,\acute{v}^{N-1}\,e^{-\acute{v}} \,\mathrm{d}\acute{v} \label{equ:CDFforUoverV} \\
f_{X}(x)&=& \frac{\mathrm{d}}{\mathrm{d}x}\,F_{X}(x)\,=\,\frac{1}{2^{2N}\,\Gamma^2(N)} \int_{0}^{\infty}\,\frac{\mathrm{d}}{\mathrm{d}x}\,\Bigg[\int_{0}^{x\acute{v}\,\sigma _v^2/\sigma _u^2} \acute{u}^{N-1}\,e^{-\acute{u}}\,\mathrm{d}\acute{u}\,\Bigg]\,\acute{v}^{N-1}\,e^{-\acute{v}} \,\mathrm{d}\acute{v} \nonumber \\
&=& \frac{1}{2^{2N}\,\Gamma^2(N)} \int_{0}^{\infty}\,\Bigg[ \bigg(\frac{\acute{v}\,\sigma _v^2}{\sigma _u^2}\bigg)\,\bigg(\frac{x\acute{v}\,\sigma _v^2}{\sigma _u^2}\bigg)^{N-1}\,e^{-x\acute{v}\,\sigma _v^2/\sigma _u^2}\,\Bigg]\,\acute{v}^{N-1}\,e^{-\acute{v}} \,\mathrm{d}\acute{v} \nonumber \\
&=& \frac{x^{N-1}}{\Gamma^2(N)}\, \bigg(\frac{\sigma _v^2}{\sigma _u^2}\bigg)^N \frac{\Gamma(2N)}{\Big(1+\sigma _v^2\,x/\sigma _u^2\Big)^{2N}}\Bigg[\int_{0}^{\infty} \frac{\Big(\acute{v}\big(1+\sigma _v^2\,x/\sigma _u^2\big)\Big)^{2N}}{2^{2N}\,\Gamma(2N)} e^{-\acute{v}\big(1+\sigma _v^2\,x/\sigma _u^2\big)} \,\frac{\mathrm{d}\acute{v}}{\acute{v}}\Bigg] \nonumber \\
&=&\frac{\sigma _v^2}{\sigma _u^2}\, \frac{\Gamma(2N)}{\Gamma^2(N)}\,\frac{\Big(\sigma _v^2\,x/\sigma _u^2\Big)^{N-1}}{\Big(1+\sigma _v^2\,x/\sigma _u^2\Big)^{2N}} \;, \;\;\;\;\;\;\;\;\;\;\;\;\; x\geq 0\label{equ:PDFforUoverV} \\
F_{X}(x)&=& Prob\big[X\leq x\big]\,=\, \int_{-\infty}^{x} f_{X}(t) \,\mathrm{d}t \,=\, \frac{\sigma _v^2}{\sigma _u^2}\, \frac{\Gamma(2N)}{\Gamma^2(N)}\,\int_{0}^{x} \frac{\Big(\sigma _v^2\,t/\sigma _u^2\Big)^{N-1}}{\Big(1+\sigma _v^2\,t/\sigma _u^2\Big)^{2N}} \,\mathrm{d}t \nonumber \\
&=& \frac{\Gamma(2N)}{\Gamma^2(N)}\,\int_{0}^{(\sigma _v^2\,x/\sigma _u^2)/(1+\sigma _v^2\,x/\sigma _u^2)} \bigg(\frac{u}{1-u}\bigg)^{N-1}\,\bigg(1+\frac{u}{1-u}\bigg)^{-2N} \,\frac{\mathrm{d}u}{\big(1-u\big)^2} \;\;\;\;\;\;\;\; \nonumber \\
&=& \frac{\Gamma(2N)}{\Gamma^2(N)}\,\int_{0}^{(\sigma _1^2\,x/\sigma _u^2)/(1+\sigma _v^2\,x/\sigma _u^2)} u^{N-1}\,(1-u)^{N-1}\,\mathrm{d}u \,=\, I_{\frac{(\sigma _v^2\,x/\sigma _u^2)}{(1+\sigma _v^2\,x/\sigma _u^2)}}(N,N) \qed \;\;\;\;\;\; \label{equ:CDFforUoverVFinal}
\end{eqnarray}
\setcounter{equation}{2}
\hrulefill
\vspace*{4pt}
\end{figure*}

\section{Energy Ratio Analysis For AWGN Channels}
\label{sec:ERanalysis}

To verify the algorithm, we first analyze the \emph{energy ratio} technique assuming perfect synchronization and neglecting the leakage power effect. However, these issues will be considered and their effects will be studied in the next section. Throughout the analysis, we assume that the signal to be detected does not have any known structure that could be exploited. Therefore, the reserved tone sequence is modelled via a zero-mean circularly symmetric complex Gaussian distribution (this is also true in case of frequency selective fading channels as discussed in section~\ref{sec:channelMIMO}). The target of this analysis is to find the receiver operating characteristics (ROC) represented by the probability of detection, $P_D$, and probability of false alarm, $P_{FA}$. The detection probability is the probability of detecting a primary signal when it is truly present while the false alarm probability is the probability that the test incorrectly decides that the primary user is present when it is actually not.

Since we are dealing with a two state model in which the channel is assumed to be idle or busy by the primary user, then we wish to discriminate between the two hypotheses $\mathcal{H}_0$ and $\mathcal{H}_1$ where the first assumes that the primary signal is not in band and the second assumes that the primary user is present. Using the \emph{energy ratio} algorithm, one can define these hypotheses as given by (\ref{equ:BinaryHypotheses}) where it is assumed that the samples contained in the first window have a variance of $\sigma_v ^2$ and the samples enclosed by the second window have a variance of $\sigma_u ^2$.
\begin{eqnarray}
&& \left\{
  \begin{array}{l l}
    \mathcal{H}_0:\;X\,=\,\frac{U}{V},  & \sigma^2_u\,=\,\sigma^2_v \\
    \mathcal{H}_1:\;X\,=\,\frac{U}{V}, & \sigma^2_u\,>\,\sigma^2_v\\
  \end{array} \right.
  \label{equ:BinaryHypotheses}
\end{eqnarray}

The performance of the detector is quantified in terms of its ROC curve, which represents the probability of detection as a function of the probability of false alarm. By varying a certain threshold $\gamma$, the operating point of a detector can be chosen anywhere along the ROC curve. $P_{FA}$ and $P_D$ can be defined as given by (\ref{equ:PFAdefinition}) and  (\ref{equ:PDdefinition}), respectively.
\begin{eqnarray}
P_{FA}&=& Prob\big[X> \gamma \,|\,\mathcal{H}_0 \big] \label{equ:PFAdefinition} \\
P_{D}&=& Prob\big[X> \gamma \,|\,\mathcal{H}_1 \big] \label{equ:PDdefinition}
\end{eqnarray}
Clearly, the fundamental problem of detector design is to choose the detection criteria, and to set the decision threshold $\gamma$ to achieve good detection performance. Detection algorithms are either designed in the framework of classical statistics, or in the framework of Bayesian statistics~\cite{ref:Detectionbook}. In the classical case, either $\mathcal{H}_0$ or $\mathcal{H}_1$ is deterministically true, and the objective is to maximize $P_D$ subject to a constraint on $P_{FA}$; this is known as the Neyman-Pearson (NP) criterion. In the Bayesian framework, by contrast, it is assumed that the source selects the true hypothesis at random, according to some priori probabilities. The objective is to minimize the so-called Bayesian cost. In this work, the former approach is followed. First, the Probability Density Function (PDF) and the Cumulative Distribution Function (CDF) of the decision variable are derived. Next, both the detection and the false alarm probabilities are evaluated in closed-forms.

\subsection{Energy Ratio PDF and CDF Evaluation}
Since the samples of the reserved tone sequence follow a zero-mean circularly symmetric complex Gaussian distribution, then the energy contained in one window will follow a Chi-Square distribution and the PDFs for the random variables $U$ and $V$ can be written as given by (\ref{equ:EnergydistributionforU}) and (\ref{equ:EnergydistributionforV}), respectively~\cite{ref:energypdf}.
\begin{eqnarray}
f_{U}(u)&=& \frac{1}{2^N\,\sigma_u ^{2N}\,\Gamma(N)}\,u^{N-1}\,e^{-u/(2\sigma_u ^2)}\, ,\;\; u>0 \label{equ:EnergydistributionforU} \\
f_{V}(v)&=& \frac{1}{2^N\,\sigma_v ^{2N}\,\Gamma(N)}\,v^{N-1}\,e^{-v/(2\sigma_v ^2)}\, ,\;\; v>0 \label{equ:EnergydistributionforV}
\end{eqnarray}

The CDF for the random variable $X$ and hence the PDF, can be evaluated as given by (\ref{equ:CDFforUoverV}) and (\ref{equ:PDFforUoverV}), respectively, where the two random variables $U$ and $V$ are assumed to be independent. It is obvious that the PDF for $X$ follows a scaled F-distribution with mean $m_X\,=\big(\Gamma(N-1)\,\Gamma(N+1)/\Gamma^2(N)\big)\times \big(\sigma^2_u/\sigma^2_v\big)$ and variance $Var(X)\,=\,\big(\Gamma(N-2)\,\Gamma(N+2)/\Gamma^2(N)\big)\times \big(\sigma^2_u/\sigma^2_v\big)^2$. The CDF for $X$ can be derived in a closed-form as given by (\ref{equ:CDFforUoverVFinal}), where $I_b(N,N)$ is the regularized incomplete beta function with the parameters $b$ and $N$.

\subsection{$P_{FA}$ and $P_D$ Evaluation}
To obtain the ROC, we develop the classical NP criterion in which the detection probability is maximized while the false alarm probability is maintained at a fixed value. Since the probability of false alarm for the \emph{energy ratio} algorithm is given by (\ref{equ:DSWProbFalseAlarm}), one can obtain the threshold $\gamma$ subjected to a constant $P_{FA}$ as given by (\ref{equ:DSWTHFromProbFalseAlarm}) where $I_b^{-1}(N,N)$ is the inverse incomplete beta function with parameters $b$ and $N$.
\begin{eqnarray}
\setcounter{equation}{11}
P_{FA}&=& Prob\big[X> \gamma \,|\,\mathcal{H}_0 \big]\,=\,1-I_{\frac{(\gamma)}{(1+\gamma)}}(N,N) \;\;\;\; \label{equ:DSWProbFalseAlarm} \\
\gamma &=& \frac{I_{1-P_{FA}}^{-1}(N,N)}{1-I_{1-P_{FA}}^{-1}(N,N)} \label{equ:DSWTHFromProbFalseAlarm}
\end{eqnarray}
Once the primary user becomes available in the band, the second window will contain the power of the primary user in addition to the power of the noise whereas the first window will contain only noise and hence, the receiver noise variance is represented by $\sigma ^2_v$. Therefore, $\sigma ^2_u\,=\,\sigma ^2_v+\mathrm{PNR}\times \sigma ^2_v$ where PNR is the ratio of the primary user power to the secondary user noise power at the secondary user receiver. Hence, the detection probability can be expressed in terms of PNR as,
\begin{eqnarray}
P_{D}&=& Prob\big[X> \gamma \,|\,\mathcal{H}_1 \big]\,=\,1-I_{\frac{(\sigma _v^2\,\gamma /\sigma _u^2)}{(1+\sigma _v^2\, \gamma /\sigma _u^2)}}(N,N) \nonumber \\
&=&1-I_{\frac{(\gamma /(1+PNR))}{(1+\gamma /(1+PNR))}}(N,N) \qed \;\;\;\;\;\; \label{equ:DSWProbDetection}
\end{eqnarray}

\section{OFDM Challenges on Energy Ratio Algorithm}
\label{sec:OFDMchallenges}

In this section we first present an overview of the current challenges faced by conventional OFDM systems and possible techniques that have been introduced to address these challenges. Finally, we show that by adopting any of these techniques, our \emph{energy ratio} detector does not require any additional complexity to the OFDM system with efficient detection capabilities.

\subsection{NBI and Power Leakage}

By definition, the power of a NBI is concentrated in a small frequency band compared to the overall system bandwidth~\cite{ref:NBIwindow}. Although the total power of the interference may be substantially lower than the total received signal power, these disturbances can reach a noise level which exceeds the received signal level by orders of magnitude inside the interference band. Therefore, the system performance will be severely degraded. Aside from NBI, the side-lobes of modulated OFDM sub-carriers even in case of having no NBI are known to be large. As a result, there is power leakage from sub-carriers to adjacent sub-carriers. It is known that the most efficient solution to NBI is to disable the sub-carriers corresponding to this interference. This will eliminate the effect of NBI at those sub-carriers, however, the signal to noise ratio at the other sub-carriers will be slightly reduced.

For the power leakage, recent research has carefully addressed this problem. For example, the out of band leakages can be reduced by including special cancelling carriers at the edge of the band~\cite{ref:outbandredusingCC}. These sub-carriers are modulated with complex weighting factors which are optimized such that the side-lobes of the those carriers cancel the side-lobes of the original transmitted signal in a certain optimization range. Another solution is proposed in~\cite{ref:precodingOFDM} where the power leakage is totally suppressed by a pre-coding technique. This pre-coding is applied to the frequency domain OFDM signal before IDFT block at the transmitter side. At the receiver, a decoder is applied to omit the spectral distortion to the OFDM signal caused by pre-coding. This technique can totally eliminate the effect of spectral leakage but of course it needs full revision for all synchronization algorithms applied to traditional OFDM system.

By utilizing the fact that \emph{energy ratio} can perfectly counter any consistent noise like signals, windowing can be applied to the time domain OFDM symbols~\cite{ref:RxWindowdesign} to limit the leakages and to reduce the influence of NBI. Thus, if a windowing function (e.g., Nyquist window) is carefully chosen to only affect the interference while leaving the OFDM signal unchanged, then spectral leakage can be avoided. In~\cite{ref:RxWindowdesign}, a folding technique is proposed in order not to use a double length DFT. In this case, the samples preceding the OFDM symbol to the end of the symbol are added to the samples following the symbol to its beginning. To evaluate the performance of our \emph{energy ratio} detector in the presence of NBI and power leakage, we turn off the sub-carriers corresponding to the NBI. Moreover, the time domain windowing technique with folding is applied at the receiver side, as it offers the lowest computational complexity with sufficiently good performance.

\subsection{Inter-carrier Interference Effect}
In addition to the NBI and power leakage problems discussed earlier, OFDM systems also suffer from ICI effects. The main sources for ICI in OFDM-based systems are the phase noise (PN), the carrier frequency offset (CFO), and the sampling frequency offset (SFO)~\cite{ref:CFOPNeffect}.

\begin{figure*}[!t]
\normalsize
\setcounter{equation}{20}
\begin{eqnarray}
f_{Re\{Y(r^k_j)\}|\,X_p(r^k_j)=a+jb}(w)=f_{Im\{Y(r^k_j)\}|\,X_p(r^k_j)=a+jb}(w)&=&\frac{1}{\sqrt{2\pi (\sigma_n^2+ E_{ab} \sigma_H^2)}}\,exp\bigg(\frac{-w^2}{2(\sigma_n^2+ E_{ab} \sigma_H^2)}\bigg) \label{equ:conditionalPDF}
\end{eqnarray}
\setcounter{equation}{17}
\hrulefill
\vspace*{4pt}
\end{figure*}

\begin{figure*}[t]
\centering\includegraphics[scale=0.38]{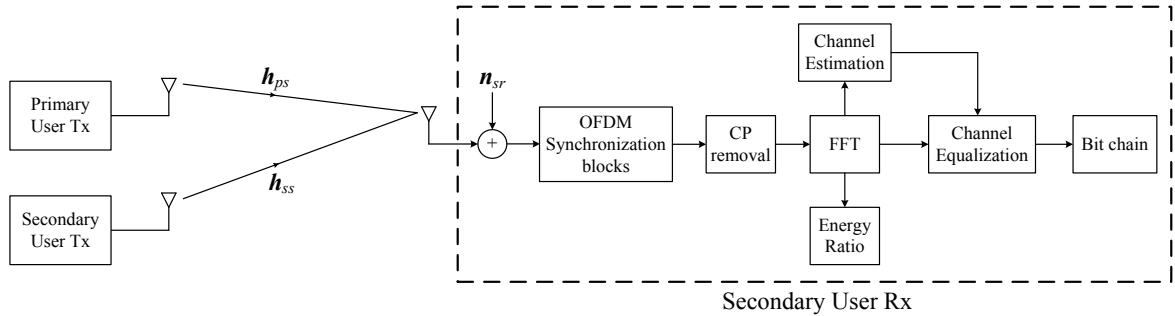}
\caption{Communication model for SISO system where primary user channel and secondary user channel are considered}
\label{fig:EnergyRatioOverFadingChannels}
\end{figure*}
The phase noise is due to the instability of carrier signal generators used at the transmitter and receiver. In fact, the effect of PN can be greatly reduced by increasing the sub-carrier spacing, $\Delta f$. We will (optimistically) assume that $\Delta f$ is large enough such that the ICI introduced by the PN is neglected with respect to the ICI generated by either CFO or SFO. On the other hand, the carrier frequency offset is due to the difference between the carrier frequencies generated by the transmitter and receiver oscillators, or by the Doppler frequency shift. It is commonly represented by the normalized CFO which is the ratio of the frequency offset to the sub-carrier spacing, defined as $\varepsilon \,=\,\varepsilon_i + \varepsilon_f$ where $\varepsilon_i$ is the integer part of the normalized CFO while $\varepsilon_f$ is the fractional part. Even after estimating and compensating both integer and fractional CFO, a residual CFO, $\varepsilon_r$, which represents the remaining uncompensated fractional CFO always exists.

For the sampling frequency offset, it is mainly caused by the mismatch between the transmitter and receiver oscillators such that the received continuous-time waveform is sampled at an interval of $(1+\delta)\,T_s$ instead of $T_s$ where $T_s$ is the ideal sampling period and $\delta$ (usually expressed in part per million or ppm) is the normalized difference between the periods of the two clocks. In~\cite{ref:JWLSCFOSFO}, $\delta$ is estimated by the receiver where compensation is carried out by feeding the clock generator with the amount of time shift in order to adjust the clock or by interpolating the received time domain samples with a fractional delay.

It is known that the residual CFO and SFO results in ICI which degrades the Signal-to-Noise Ratio (SNR) over all sub-carriers. The SNR degradation, $\mathrm{SNRD}_\mathrm{CFO}$, due to the residual CFO is analytically analyzed in~\cite{ref:carrierdistortion}. The analysis shows that in the Additive White Gaussian Noise (AWGN) channel and when the number of sub-carriers is large, the SNR degradation is given by,
\begin{eqnarray}
\mathrm{SNRD}_\mathrm{CFO}\bigg|_{\mathrm{dB}}&=&\frac{10}{3\,\mathrm{ln}(10)} \big(\pi \, \varepsilon _r\big)^2\,  \mathrm{SNR} \label{equ:SNRDegDueToCFO}
\end{eqnarray}
Similarly, the SNR degradation due to the residual SFO, $\delta_r$,
in the $k^{\mathrm{th}}$ sub-carrier,
$\mathrm{SNRD}_\mathrm{SFO}(k)$, is analyzed
in~\cite{ref:SNRDSFOEffect} and is given by,
\begin{eqnarray}
\mathrm{SNRD}_\mathrm{SFO}(k)\bigg|_{\mathrm{dB}}&=&10\,\mathrm{log}_{\mathrm{10}} \bigg(1+\frac{1}{3} \big(\pi\,\delta_r \,k\big)^2\, \mathrm{SNR} \bigg)\;\;\;\;\;\; \label{equ:SNRDegDueToSFO}
\end{eqnarray}

Since CFO and SFO estimation and compensation is a must for traditional OFDM systems, we have to consider these issues when the \emph{energy ratio} algorithm is evaluated in the presence of ICI. Our goal here is to emphasize that the \emph{energy ratio} technique does not require any new solutions for the OFDM synchronization problems. Indeed, the \emph{energy ratio} can provide a very good performance even with the existing algorithms for the OFDM synchronization engine.

\subsubsection{CFO Estimation and Compensation}

Any practical system assumes a maximum acceptable frequency offset, $\mathrm{CFO_{max}}$, between the transmitter and receiver. Therefore, the integer CFO range is known by the maximum integer CFO, ${\varepsilon_i}_{\mathrm{max}}=\lfloor \mathrm{CFO_{max}}/\Delta f\rfloor$. Hence, the integer CFO range will be $\mathbb{L}=\big[-{\varepsilon_i}_{\mathrm{max}}\;-{\varepsilon_i}_{\mathrm{max}}+1\;...\;-1\;0\;1\;...\;{\varepsilon_i}_{\mathrm{max}}-1\;{\varepsilon_i}_{\mathrm{max}}\big]$. In~\cite{ref:OFDMsynch}, a two step time domain estimation technique is introduced for CFO. This approach depends on the training symbols that are transmitted at the front of the OFDM frame. Actually, a good compromise between performance and complexity is achieved by this technique. The idea is to first estimate the fractional CFO by a maximum likelihood estimator as given by (\ref{equ:CFOFractionalest}) where $y(n)$ is the received time domain signal and $D\,=\,N_s+N_g$.
\begin{eqnarray}
\hat{\varepsilon}_f&=&\frac{1}{2\pi D} \angle \Bigg\{ \sum_{n=0}^{n=N_s-1} y(n)\,y^*(n+D)\Bigg\} \label{equ:CFOFractionalest}
\end{eqnarray}
It applies an autocorrelation to the time domain waveform with the condition that two or more training symbols are inserted at the beginning of the frame. The time domain signal is compensated for the fractional CFO resulting in the signal $y_{\mathrm{comp}}(n)$. This signal is then cross-correlated with the transmitted time domain waveform for the training symbols, $y_t$,  after applying a progressive phase shift that depends on the desired integer CFO as given by (\ref{equ:CFOIntest}).
\begin{eqnarray}
\hat{\varepsilon}_i&=&\max_{m\in \mathbb{L}} \Bigg| \sum_{n=0}^{n=N_s-1} y_{\mathrm{comp}}(n)\,y_t^*(n)\,e^{-2\pi j m n/D} \Bigg| \label{equ:CFOIntest}
\end{eqnarray}
This cross-correlation is repeated for each integer CFO in $\mathbb{L}$ and the maximum is searched for. The integer CFO that corresponds to the maximum correlation is selected as the estimated integer CFO. Once the normalized CFO is estimated, the OFDM signal can be compensated by rotating the phase of the time domain signal by $-2\pi(\hat{\varepsilon}_f+\hat{\varepsilon}_i) n$ where $n$ is the time index.

\subsubsection{SFO Estimation and Compensation}
\label{sec:SFOEstCOmp}
In~\cite{ref:JWLSCFOSFO}, the carrier-frequency and timing offsets are jointly estimated by applying a weighted least-squares (WLS) algorithm where a weighting matrix, $\mathbf{W}$, is designed to improve the estimation accuracy of the least-squares. The analytical results in~\cite{ref:JWLSCFOSFO} show that this matrix should be a function of the noise variance. In fact, if an incorrect (estimated) value of the noise variance is used, then the resulting estimation accuracy may perform rather poorly. Since the \emph{energy ratio} is strong enough to compact ICI, we can simply apply the WLS algorithm by replacing $\mathbf{W}$ with an identity matrix. This reduces the WLS algorithm into the well-known least-squares estimation. First, we compute the averaged phase difference between the pilots contained in two consecutive OFDM training symbols in the frequency domain to obtain $\mathbf{y}=\big[\mathit{y}_0\;\mathit{y}_1\;...\;\mathit{y}_{\mathit{J}-1}\big]^T$ where $\mathit{J}$ is the number of pilots inserted in one preamble symbol. Second, the pilot sub-carrier indices denoted by $\mathit{x_j}, j\,=\,0,\,1,\,2,\,...\,\mathit{J}-1$ are arranged to construct the matrix $\mathbf{X}$ which is given by (\ref{equ:SFOXmatrix}). Finally, the estimated carrier-frequency offset $\hat{\varepsilon}$ and timing offset $\hat{\delta}$ can be obtained by (\ref{equ:SFOlinearsystemsol}).
\begin{eqnarray}
\mathbf{X}&=&\left[ \begin{array}{ccccc}
\mathit{x}_0&\mathit{x}_1&\mathit{x}_2&\ldots& \mathit{x}_{\mathit{J}-1} \\
1&1&1&\ldots& 1
\end{array} \right]^T \label{equ:SFOXmatrix} \\
\Big[\hat{\delta}\;\;\;\hat{\varepsilon}\Big]^T&=&\frac{N_s}{2\pi (N_s+N_g)}\big(\mathbf{X}^*\;\mathbf{X}\big)^{-1}\;\mathbf{X}^*\;\mathbf{y} \label{equ:SFOlinearsystemsol}
\end{eqnarray}

\section{Frequency Selective Channel And Multi-Antenna System}
\label{sec:channelMIMO}
To study the effect of the frequency selective fading channel on the \emph{energy ratio} algorithm, we first consider the single-input single output (SISO) model shown in Fig.~\ref{fig:EnergyRatioOverFadingChannels} where the secondary transmitter communicates with one SU slave over the channel $h_{ss}$. During the transmission, the primary user may attempt transmission which is received by the secondary receiver across the channel $h_{ps}$. Both signals are combined at the receiver antenna and then processed as one received stream. The receiver noise is added to the combined signals and the result is converted to the frequency domain by the DFT block. The reserved tone sequence is then organized in order to be processed by the monitoring algorithm.

If $r^k_i, i=0,\, 1,..., N_{RT}-1$ denotes the reserved tone indices for the $k^{\mathrm{th}}$ OFDM symbol, then the $j^{\mathrm{th}}$ reserved tone can be modelled as given by (\ref{equ:ReservedTonesValues}) where $X_s(r^k_j)$, $X_p(r^k_j)$, $H_{ss}(r^k_j)$, $H_{sp}(r^k_j)$, and $n(r^k_j)$ are the secondary user transmitted symbol, the primary user transmitted symbol, the frequency domain response for the secondary channel, the frequency domain response for the primary channel, and the noise sample, respectively, where all are observed at sub-carrier $r^k_j$. Indeed, this is one of the most important properties for OFDM technique in which the frequency selective fading can be converted into flat fading over each sub-carrier. Since the secondary transmitter forces the reserved tones to be null, then $X_s(r^k_j)= 0, \; \forall j$ and hence the received reserved tones include the effect of the primary user and the noise of the secondary receiver under perfect synchronization and neglecting the power leakage effect.
\begin{eqnarray}
Y(r^k_j)&=&H_{ps}(r^k_j)X_p(r^k_j)+H_{ss}(r^k_j)X_s(r^k_j)+n(r^k_j) \nonumber \\
&=&H_{ps}(r^k_j)X_p(r^k_j)+n(r^k_j) \label{equ:ReservedTonesValues}
\end{eqnarray}

Now, suppose that the primary-to-secondary channel impulse response, $h_{ps}$, is modelled by a finite impulse response (FIR) filter with $N_g$ taps where each tap $l$ has the channel gain $h_{ps}(l)$ for $l\,= \,0,\,1, ... ,\, N_g-1$. Here, we assume that the maximum delay of the multi-path fading channel is fully characterized by the cyclic prefix length. If we denote $\sigma_H^2$ as the sum of the channel tap powers such that $\sigma_H^2\,=\,\sum_{l=0}^{N_g-1} E\big[|h_{ps}(l)|^2\big]$, then the conditional probability density function for either the real part or the imaginary part of the received symbol at index $r^k_j$ given that the transmitted symbol is $X_p(r^k_j)\,=a+jb$ can be obtained by (\ref{equ:conditionalPDF}) where $\sigma_n^2$ is the noise variance and $E_{ab}\,=\,(a^2+b^2)$~\cite{ref:OFDMsignalPDF}.

If the PU uses any Phase Shift Keying (PSK) modulation like QPSK, then the average symbol energy is simply $E_s\,=\,E_{ab}\,=a^2+b^2$ and the received signal is modelled by a circularly symmetric Gaussian distribution with zero-mean and variance $\sigma_n^2+ E_s \sigma_H^2$. In this case, the \emph{energy ratio} algorithm can still detect the reappearance of the primary user when the first window is filled with unwanted signals (i.e: $\sigma_v^2\,=\,\sigma_n^2$) and the second window includes both the unwanted signals and the primary user signal (i.e: $\sigma_u^2\,=\,\sigma_n^2+ E_s \sigma_H^2$). The same performance as the AWGN case can be obtained. However, the primary to secondary power ratio is defined as $PNR\,=\,E_s /\sigma_n^2$ and hence the probability of detection will depend on the channel profile as given by (\ref{equ:DSWProbDetection2}) where $\sigma_u^2\,=\,\sigma_v^2\,(1+ PNR\times \sigma_H^2)$. The conclusion is that the \emph{energy ratio} algorithm can behave as in AWGN channel even when the channel is frequency-selective for both primary and secondary users.
\setcounter{equation}{21}
\begin{eqnarray}
P_{D}&=& Prob\big[X> \gamma \,|\,\mathcal{H}_1 \big]\,=\,1-I_{\frac{(\sigma _v^2\,\gamma /\sigma _u^2)}{(1+\sigma _v^2\, \gamma /\sigma _u^2)}}(N,N) \nonumber \\
&=&1-I_{\frac{(\gamma /(1+\sigma_H^2 PNR))}{(1+\gamma /(1+\sigma_H^2 PNR))}}(N,N)  \label{equ:DSWProbDetection2}
\end{eqnarray}

To enhance the detector performance in fading channels, multiple-antennas at the receiver side can be utilized. For Single-Input Multiple-Output (SIMO) or Multiple-Input Multiple-Output (MIMO) systems, if the number of receive antennas is $N_{Rx}$, there will be $N_{Rx}$ available sets of reserved tones at the receiver for each OFDM symbol or equivalently $N_{Rx}\times N_{RT}$ reserved tones every OFDM symbol. The \emph{energy ratio} monitoring technique will combine all these sets to form the reserved tone sequence. In this case, the confidence of primary user presence is increased by the diversity gain offered by the system. This allows for more robust decision compared to the SISO case. Effectively, applying SIMO or MIMO is equivalent to increasing the window size by a factor of $N_{Rx}$. If the same performance is required, the window size can be reduced by $N_{Rx}$ which implies that the primary user power is sensed in less time when compared to the SISO case. Otherwise, increasing the window size directly increases the mean of the decision making variable under $\mathcal{H}_1$ which allows for higher detection probability and less false alarm.

\section{Complexity Overhead For Energy Ratio Algorithm}
\label{sec:complexity}
\begin{figure}[t]
\centering\includegraphics[scale=0.31]{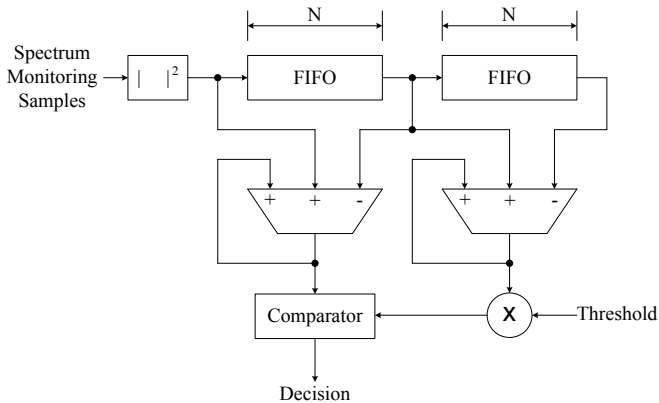}
\caption{Proposed architecture for the energy ratio algorithm}
\label{fig:DSWImplementation}
\end{figure}
To evaluate the \emph{energy ratio} from complexity point of view, we propose an architecture for the algorithm and then analyze the corresponding complexity and compare it to the traditional energy detectors. The proposed architecture is shown in Fig.~\ref{fig:DSWImplementation}. First, the reserved tone sequence is injected to be squared. Next, two First-In First-Out (FIFO) memories are used to store the squared outputs in order to manage the energy evaluation for the two windows. The idea depends on the sliding concept for the windows where the total energy enclosed by one window can be evaluated by only adding the absolute squared of the new sample and subtracting the absolute squared of the last sample in the previous window as given by,
\begin{eqnarray}
V(k)&=&\sum_{i=k}^{N+k-1} \big| Z_i \big| ^2 \nonumber \\
&=&V(k-1)+ \big| Z_{N+k-1} \big| ^2 -\big| Z_{k-1} \big| ^2\label{equ:EnergyExpressionSlidingWindow}
\end{eqnarray}
The ratio may not be evaluated directly, instead we can multiply the energy of the first window by the threshold and the multiplication output is then compared to the energy of the second window. We conclude that the proposed architecture typically uses double the components applied for the traditional energy detector. Moreover, traditional spectrum sensing which is applied prior to spectrum monitoring surely involve multipliers and accumulators. To further reduce the complexity, these modules can be reused and shared with the \emph{energy ratio} algorithm during spectrum monitoring as sensing and monitoring are non-overlapped in time.
\begin{figure}[t]
\centering\includegraphics[scale=0.46]{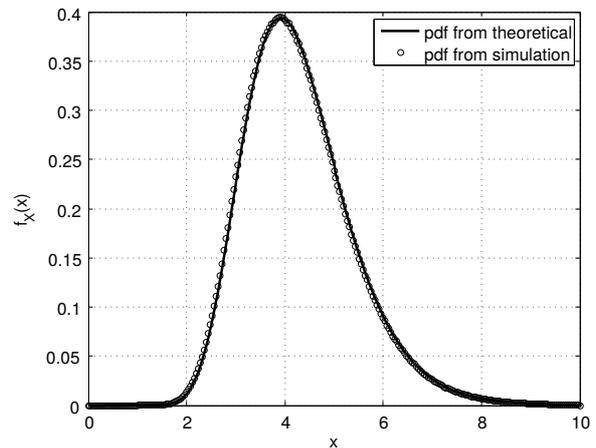}
\caption{Simulated PDF versus analytical PDF for the energy ratio decision making variable with $N$ =32 and $10\,\mathrm{log_{10}}(\sigma _u^2/\sigma _v^2)\,=$ 5dB}
\label{fig:pdfSimVsTh2}
\end{figure}

\section{Simulation Results}
\label{sec:Results}
In the simulation, we used an OFDM system that employs a total of $N_s\,=$ 1024 sub-carriers, 224 of which are used as guard bands on both ends of the signal band. There are 32 pilot sub-carriers and $N_{RT}\,=$ 4 reserved tones, distributed across the entire 800 sub-carriers. Therefore, the throughput reduction due to reserved tones is only $0.5\%$ which is an inconsiderable amount for high data rates. The cyclic prefix is $N_g\,=$ 64 samples long and the sampling frequency is 16MHz. The sub-carrier spacing is then $\Delta f\,=\,15.625$ KHz which is large enough to neglect the phase noise distortion and the time domain windowing effect. Unless otherwise specified, the frame has two consecutive training symbols, 256 OFDM data symbols, and the reserved tone spacing $\Delta_r\,=2$. The data for both primary and secondary transmitters is modulated by 16-PSK mapper and the secondary power to noise ratio in the absence of primary signal is assumed 9dB. When the system operates under non-perfect synchronization, the maximum acceptable CFO is assumed to be 400KHz, the CFO is 320KHz, and the sampling clock offset is assumed to be 100 ppm.
\begin{figure}[t]
\centering\includegraphics[scale=0.48]{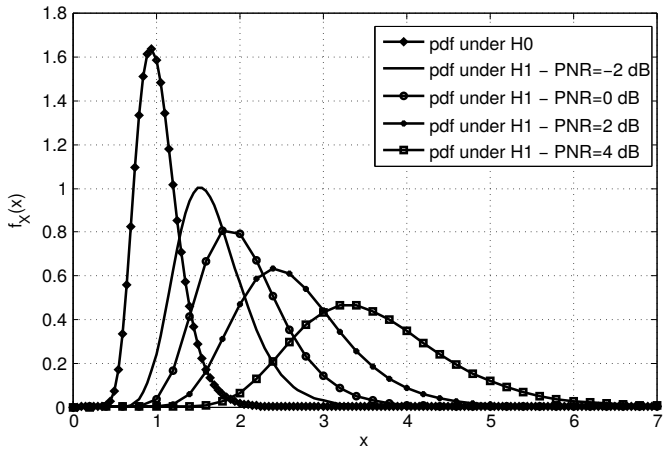}
\caption{Conditional PDF under $\mathcal{H}_0$ and conditional PDF under $\mathcal{H}_1$ for PNR=-2, 0, 2, and 4 dB}
\label{fig:PDFH0H12}
\end{figure}
\subsection{Analytical Verification}
Fig.~\ref{fig:pdfSimVsTh2} shows a comparison between the PDF given by (\ref{equ:PDFforUoverV}) and the one obtained from simulation where we have used $10\,\mathrm{log_{10}}(\sigma _u^2/\sigma _v^2)\,=$ 5dB and an energy ratio window $N\,=$ 32. To obtain the simulated PDF, $10^7$ circularly symmetric Gaussian distributed samples are generated and scaled properly for both windows. The samples are then applied to the \emph{energy ratio} algorithm and the PDF is obtained by considering the histogram of the decision making variable. It is obvious that the analytical results are in excellent agreement with the simulated ones.

Next, the hypothesis test is to be verified by exploring the conditional PDF under both $\mathcal{H}_0$ and $\mathcal{H}_1$. In fact, when there is no primary user in band, the decision variable follows only one unique PDF that is shown in Fig.~\ref{fig:PDFH0H12}. Under $\mathcal{H}_1$, the conditional PDF depends on the PNR ratio. Four additional curves are also shown in Fig.~\ref{fig:PDFH0H12} for the conditional PDF under $\mathcal{H}_1$ with four different PNR values (-2, 0, 2, and 4 dB). It is clear that the decision variable can distinguish between no primary user case and primary user presence based on the PNR.

\subsection{Receiver Characteristics}
\begin{figure}[t]
\centering\includegraphics[scale=0.48]{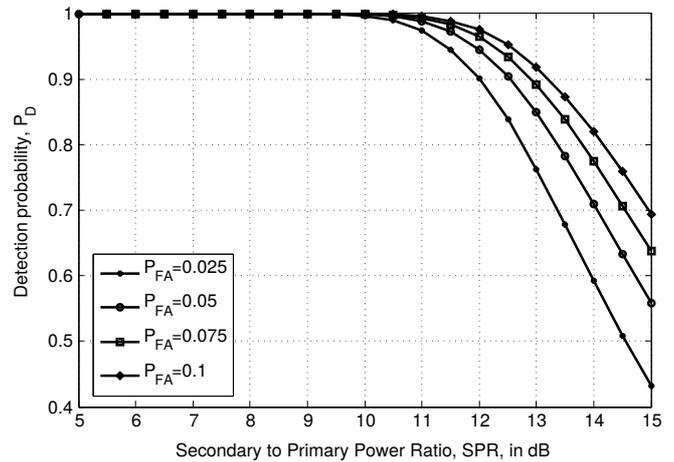}
\caption{The detection probability at fixed false alarm probability under perfect synchronization and neglecting the power leakage effect}
\label{fig:ERAWGN2}
\end{figure}

The detection probability for four different false alarm probabilities is shown in Fig.~\ref{fig:ERAWGN2}. The horizontal axis denotes the secondary to primary power ratio (SPR) which is related to the primary to secondary noise ratio (PNR) such that PNR$\mathrm{\big|_{dB}}$ = SNR$\mathrm{\big|_{dB}}$- SPR$\mathrm{\big|_{dB}}$, where SNR is the secondary power to noise power ratio. It is to be noticed that, PNR is the ratio that determines the performance of the \emph{energy ratio} algorithm whereas SPR is assumed to be the main parameter by which a monitoring algorithm is evaluated.

\begin{figure}[t]
\centering\includegraphics[scale=0.43]{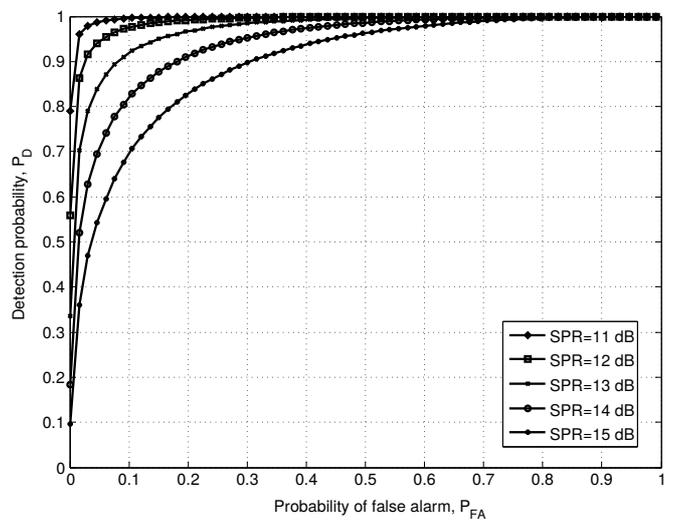}
\caption{Receiver operating characteristics for different SPR values under perfect synchronization and neglecting the power leakage effect}
\label{fig:ROCAWGN2}
\end{figure}

\begin{figure}[t]
\centering\includegraphics[scale=0.53]{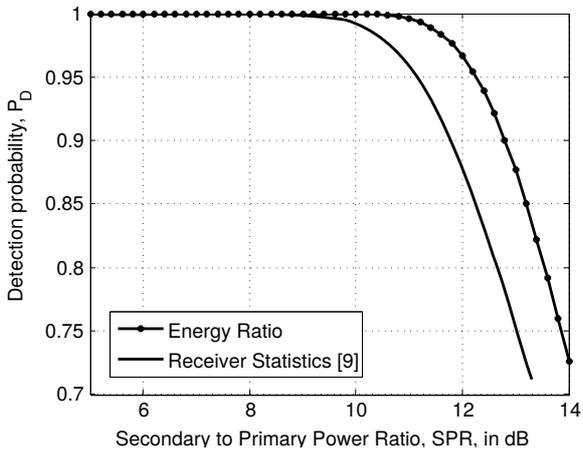}
\caption{Comparison between energy ratio and receiver statistics algorithms in case of QPSK, SNR$\,=6\,$dB, $P_{FA}\,=0.04$, and $N\,=$128}
\label{fig:RxStatvsEnergyRatio2}
\end{figure}

The ROC for the \emph{energy ratio} for different values of SPR is shown in Fig.~\ref{fig:ROCAWGN2}. These results are obtained by simulating the OFDM system twice, one when primary signal is present and the second when it is absent. The system is run over $10^6$ realizations and the probability of detection or false alarm is evaluated. The threshold is set based on the theoretical value given by (\ref{equ:DSWTHFromProbFalseAlarm}).
In order to compare the proposed monitoring algorithm with the receiver statistics technique found in~\cite{ref:MonitoringDuringReception}, the OFDM system is simulated such that the system parameters match the simulation environment followed by~\cite{ref:MonitoringDuringReception}. The simulation is run for 4-QAM under SNR$\,=\,$6dB, $P_{FA}$ =0.04, and $N$=128. Fig.~\ref{fig:RxStatvsEnergyRatio2} shows the simulation results for the detection probability of the \emph{energy ratio} algorithm in comparison with the results obtained in~\cite{ref:MonitoringDuringReception}. In addition of having fast detection, it is noted that the \emph{energy ratio} shows a better performance than the receiver statistics algorithm.
\begin{figure}[t]
\centering\includegraphics[scale=0.55]{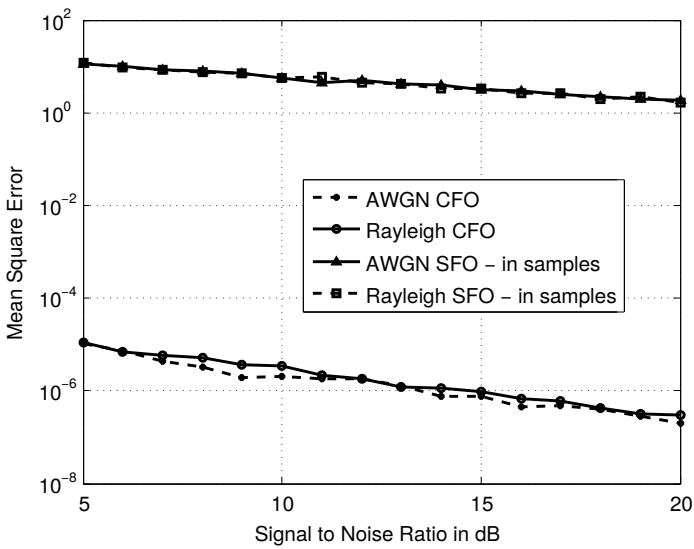}
\caption{MSE for both CFO and SFO estimation under AWGN and frequency selective fading channels. The MSE for SFO is measured in samples.}
\label{fig:CFOSFOMSE2}
\end{figure}

\subsection{OFDM Challenges}
The power leakage is modelled by applying oversampling to the frequency domain signal, where the number of points at the receiver DFT is four times the number of points used at the transmitter. Time domain Hanning window with folding is applied at the receiver to limit the NBI and power leakage. Also the phase of the time domain samples is rotated by $2\pi \varepsilon n$ to model the receiver CFO where $n$ is the time index. Moreover, the received signal is re-sampled at time instances that are multiple of $(1+\delta)T_s$ to model the receiver SFO. The preamble detection and the exact frame timing are assumed to be perfect. Here the time domain preamble is used to estimate and compensate for the CFO. The CFO compensated signal is converted to the frequency domain via DFT. The SFO, $\hat{\delta}$, and the residual CFO are further estimated by applying the least squares algorithm discussed in~\ref{sec:SFOEstCOmp}. Moreover, the time domain signal is re-sampled according to the delay $\hat{\delta}$ to compensate for the SFO.

Fig.~\ref{fig:CFOSFOMSE2} shows the mean square error for the estimated CFO and SFO. From these results, we can see that the residual fractional CFO and SFO at 9dB are $9\times 10^{-3}$ and $5\times 10^{-6}$, respectively. This implies SNR degradation of $\mathrm{SNRD}_\mathrm{CFO}\,=\,0.0092$ dB for CFO, and $\mathrm{SNRD}_\mathrm{SFO}(1023)\,=\,0.003$ dB for SFO at the last sub-carrier, based on (\ref{equ:SNRDegDueToCFO}) and (\ref{equ:SNRDegDueToSFO}), respectively. This shows the advantages of the powerful estimation techniques we have chosen for the OFDM synchronization engine.

\begin{figure}[t]
\centering\includegraphics[scale=0.47]{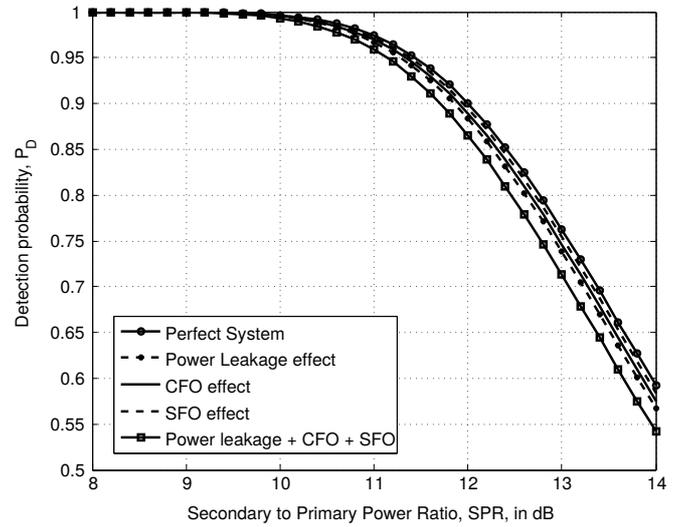}
\caption{Power leakage, CFO, and SFO effects on the energy ratio algorithm at $P_{FA}\,=0.025$}
\label{fig:synchEffects2}
\end{figure}

To examine the combined effects of OFDM impairments, the detection probability for the \emph{energy ratio} is simulated in the presence of power leakage, CFO, and SFO as shown in Fig.~\ref{fig:synchEffects2}. The signal is oversampled four times by applying 4096 points DFT at the receiver in order to allow the emulation of the continuous spectrum. The sub-carriers are then selected by sampling this spectrum every four samples. Since the sub-carrier sinc shape becomes more narrow because of the Hanning window, the introduced ICI by the residual CFO and SFO errors has a very small noticed degradation. Therefore, if windowing, CFO and SFO estimations and compensations are applied, the power leakage to neighbouring sub-carriers does not introduce severe degradation to the PU detection. As we claimed earlier, the \emph{energy ratio} is shown to be robust to OFDM challenges as only minor degradation in detection performance is noted compared to the perfect case. For instance, the overall loss due to all impairments is only 0.4 dB at a detection probability $P_D\,=0.9$.
\begin{figure}[t]
\centering\includegraphics[scale=0.54]{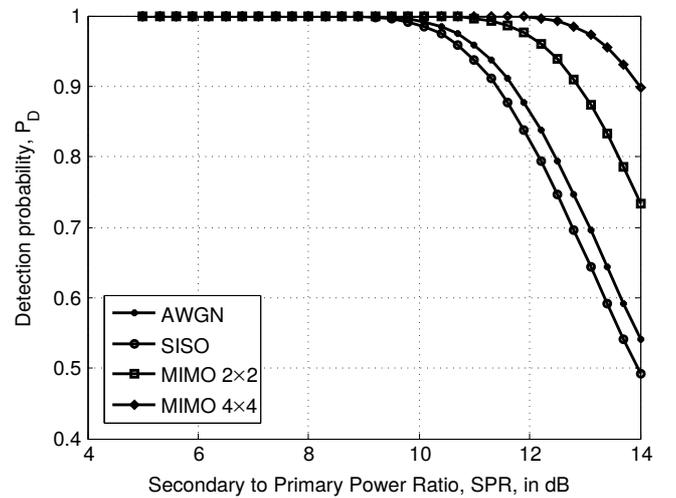}
\caption{Frequency selective fading channel effect on energy ratio for SISO and MIMO systems taking power leakage and ICI into consideration, $P_{FA}\,=0.025$ and $N\,=128$}
\label{fig:CHMIMO2}
\end{figure}

\subsection{Effect of Frequency-Selective Fading}
To study the effect of frequency-selectivity on the proposed energy ratio technique, the channel is modelled as a linear time-varying filter whose impulse response, $h(n)$, is obtained by: (1) $N_g$ circularly symmetric Gaussian samples with unit variance. The number of channel taps is defined by the cyclic prefix length as we assume that the cyclic prefix fully defines the channel maximum excess delay. (2) The samples are scaled to fit the required power delay profile which is assumed to be exponentially decaying~\cite{ref:PDPexponential}. Thus, the channel tap $l$ is scaled by $exp(-l)$ for $l\,=\,0,\,1,\,2,...,\,N_g-1$.

The OFDM system is simulated in frequency selective channel for different SPR. In Fig.~\ref{fig:CHMIMO2}, we show the effect of frequency selective fading channel on the \emph{energy ratio} performance for SISO, $2\times 2$ MIMO, and $4\times 4$ MIMO systems. The fading channel effect is compared with the AWGN channel where a minor degradation is noticed due to the narrow band problem. From these results, it is clear that having more receive antennas will offer great enhancement to the detection accuracy of the \emph{energy ratio} detector.


\section{Conclusion}
\label{sec:conclusion}

We proposed a spectrum monitoring algorithm that can sense the reappearance of the primary user during the secondary user transmission. This algorithm, named "\emph{energy ratio}" is designed for OFDM systems such as Ecma-392 and IEEE 802.11af systems. We also derived the detection probability and the probability of false alarm for AWGN channels in order to analyze the performance of the proposed algorithm. Simulation results indicate that the detection performance is superior than the receiver statistics method. For computational complexity, the \emph{energy ratio} architecture is investigated where it was shown that it requires only about double the complexity of the conventional energy detector. When frequency-selective fading is studied, the \emph{energy ratio} algorithm is shown to achieve good performance that is enhanced by involving SIMO or MIMO systems. We have proven that the multiple receive antenna system will further result in a better detection accuracy by emulating the increase in sliding window size. Therefore, our proposed spectrum monitoring algorithm can greatly enhance the performance of OFDM-based cognitive networks by improving the detection performance with a very limited reduction in the secondary network throughput.

\bibliography{References}
 \bibliographystyle{IEEEtran}

\end{document}